\documentclass[
    ,final            
  ]
  {aipproc}

\usepackage{axodraw,amsmath,rotating}

\layoutstyle{8x11double}

\graphicspath{{figs/}}

\input{paperdef}

\begin{document}

\onecolumn

\thispagestyle{empty}
\setcounter{page}{0}
\def\thefootnote{\fnsymbol{footnote}}

\begin{flushright}
\mbox{}
arXiv:0809.2395 [hep-ph]
\end{flushright}

\vspace{1cm}

\begin{center}

{\fontsize{15}{1} 
\sc {\bf How To Determine SUSY Mass Scales Now}}
\footnote{plenary talk given at the {\em SUSY\,08}, 
June 2008, Seoul, Korea}

\vspace{1cm}

{\sc 
S.~Heinemeyer%
\footnote{
email: Sven.Heinemeyer@cern.ch
}
}

\vspace*{1cm}

Instituto de Fisica de Cantabria (CSIC-UC), Santander,  Spain

\end{center}

\vspace*{0.2cm}

\BC {\bf Abstract} \EC
Currently available experimental data from electroweak precision
observables (EWPO), $B$-physics observables (BPO) and cosmological data
can be combined to extract the preferred value of SUSY mass scales. We
review recent results on the predictions of the masses of supersymmetric
particles and the indirect determination of the lightest Higgs boson
mass. Special emphasis is put on models going beyond the Constrained
Minimal Supersymmetric Standard Model (CMSSM), such as the Non-Universal
Higgs Model type~I (NUHM1), or gauge and anomaloy mediated SUSY
breaking. 

\def\thefootnote{\arabic{footnote}}
\setcounter{footnote}{0}

\newpage


{
\twocolumn

\title{How To Determine SUSY Mass Scales Now}


\author{S.~Heinemeyer}{
  address={Instituto de Fisica de Cantabria (CSIC-UC), Santander, Spain}
}



\begin{abstract}
Currently available experimental data from electroweak precision
observables (EWPO), $B$-physics observables (BPO) and cosmological data
can be combined to extract the preferred value of SUSY mass scales. We
review recent results on the predictions of the masses of supersymmetric
particles and the indirect determination of the lightest Higgs boson
mass. Special emphasis is put on models going beyond the Constrained
Minimal Supersymmetric Standard Model (CMSSM), such as the Non-Universal
Higgs Model type~I (NUHM1), or gauge and anomaloy mediated SUSY
breaking. 
\end{abstract}

\maketitle


\section{INTRODUCTION}

The Standard Model (SM) cannot be the ultimate theory of
particle physics. While describing direct experimental data reasonably
well, it fails to include gravity, it does not provide cold dark matter,
and it has no solution to the hierarchy problem, i.e.\ it does not have
an explanation for a Higgs-boson mass at the electroweak scale. 

Theories based on Supersymmetry (SUSY)~\cite{mssm} are widely
considered as the theoretically most appealing extension of the
SM. They are consistent with the approximate
unification of the gauge coupling constants at the GUT scale and
provide a way to cancel the quadratic divergences in the Higgs sector
hence stabilizing the huge hierarchy between the GUT and the Fermi
scales. Furthermore, in SUSY theories the breaking of the electroweak
symmetry is naturally induced at the Fermi scale, and the lightest
supersymmetric particle can be neutral, weakly interacting and
absolutely stable, providing therefore a natural solution for the dark
matter problem.
SUSY predicts the existence of scalar partners $\tilde{f}_L,
\tilde{f}_R$ to each SM chiral fermion, and spin--1/2 partners to the
gauge bosons and to the scalar Higgs bosons. 
The Higgs sector of the Minimal Supersymmetric Standard Model (MSSM) 
with two scalar doublets accommodates five physical Higgs bosons. In
lowest order these are the light and heavy $\cp$-even $h$
and $H$, the $\cp$-odd $A$, and the charged Higgs bosons $H^\pm$.
So far, the direct search for SUSY particles has not been successful.
One can only set lower bounds of ${\cal O}(100)$~GeV on
their masses~\cite{pdg}. 

Besides the direct detection of SUSY particles (and Higgs bosons), 
physics beyond the SM can also be probed by precision observables via the
virtual effects of the additional particles.
Observables (such as particle masses, mixing angles, asymmetries etc.)
that can be predicted within the MSSM and thus depend sensitively
on the model parameters constitute a test of the model on the
quantum level. 
The most relevant electroweak precision observables (EWPO) in this
context are the $W$~boson mass, $\MW$, the effective leptonic weak
mixing angle, $\sweff$, and the anomalous magnetic moment of the muon, 
$\amu \equiv (g-2)_\mu/2$. Since the lightest MSSM Higgs boson mass, $\Mh$
can be predicted, it also constitutes a precision observable.
An overview of SUSY effects on EWPO can be found in \citere{PomssmRep}.

An example how the EWPO can restrict the SM or the MSSM parameter space
is shown in \reffi{fig:MWSW}~\cite{FeynW,FeynZ}%
\footnote{
The plot is an update from \citere{FeynZ}.%
}%
. The plot compares the the combined prediction for $\MW$ and $\sweff$
in the SM and the MSSM. The predictions within the two models 
give rise to two bands in the $\MW$--$\sweff$ plane with only a
relatively small overlap sliver (indicated by a dark-shaded (blue) area in
\reffi{fig:MWSW}).  
The allowed parameter region in the SM (the medium-shaded (red)
and dark-shaded (blue) bands) arises from varying two parameters:
the mass of the SM Higgs boson, from $\MHSM = 114\gev$,
the LEP exclusion bound~\cite{LEPHiggsSM} to $400 \gev$ as indicated by
an arrow; the other parameter is the top quark mass, which has
been varied from $165 \gev$ to $175 \gev$, which is also
indicated by an arrow, where the current experimental value is 
$\mt^{\rm exp} = 172.4 \pm 1.2 \gev$~\cite{mt1724}, 
The light shaded (green) and the dark-shaded (blue) areas indicate 
allowed regions for the unconstrained MSSM, obtained from scattering the
relevant parameters independently~\cite{FeynZ}. 
The decoupling limit with SUSY masses of \order{2 \tev}
yields the upper edge of the dark-shaded (blue) area. Thus, the overlap 
region between
the predictions of the two models corresponds in the SM to the region
where the Higgs boson is light, i.e.\ in the MSSM allowed region 
($\Mh \lsim 135 \gev$~\cite{mhiggslong,mhiggsAEC}). In the MSSM it
corresponds to the case where all 
superpartners are heavy, i.e.\ the decoupling region of the MSSM.
The current experimental limits~\cite{LEPEWWG,TEVEWWG}, 
\begin{align}
\MW^{\rm exp} &= 80.399 \pm 0.025 \gev~, \\
\sweff^{\rm exp} &= 0.23153 \pm 0.00016
\end{align}
are indicated in the plot. As can be seen from 
\reffi{fig:MWSW}, the current experimental 68\%~C.L.\ region for 
$\MW$ and $\sweff$ exhibits a slight preference of the MSSM over the SM.

\begin{figure}[htb!]
\centerline{
\includegraphics[width=0.48\textwidth,height=5.5cm]{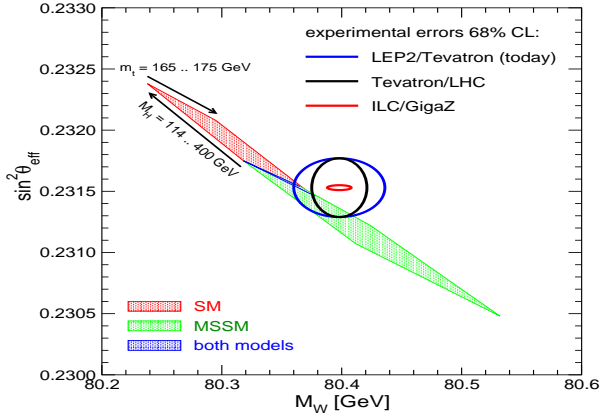}
\begin{picture}(0,0)
\CBox(-105,018)(-13,028){White}{White}
\end{picture}
}
\caption{
Prediction for $\MW$ and $\sweff$ in the MSSM and the SM (see
text)~\cite{FeynZ}.  
Also shown are the present experimental results for $\MW$ and $\sweff$
at the 68\%~C.L.
}
\label{fig:MWSW}
\end{figure}

Other important ingredients  for the determination of SUSY mass scales
from current experimental data are $B$-physics observables (BPO), see
e.g.\ \citeres{LSPlargeTB,hfag} and the abundance of cold dark matter
(CDM) in the early universe~\cite{wmap}, where
the lightest SUSY particle (LSP), assumed to be the lightest neutralino,
is required to give rise to the correct amount of cold dark matter (CDM). 

The dimensionality of the parameter space of the MSSM is so high that
phenomenological analyses often make simplifying assumptions 
that reduce drastically the number of parameters. 
One assumption that is frequently employed is
that (at least some of) the soft SUSY-breaking parameters are universal
at some high input scale, before renormalization. 
One model based on this simplification is the 
constrained MSSM (CMSSM), in which all the soft SUSY-breaking scalar
masses $m_0$ are assumed to be universal at the GUT scale, as are the
soft SUSY-breaking gaugino masses $m_{1/2}$ and trilinear couplings
$A_0$. Further parameters of this model are $\tb \equiv v_2/v_1$, 
the ratio of the two vacuum expectation values and the sign of the Higgs
mixing parameter~$\mu$.
An interesting deviation from the CMSSM is that the soft SUSY-breaking
contribution(s) to the Higgs scalar masses-squared, $m_H^2$
($m_{H_u}^2$ and $m_{H_d}^2$), is allowed 
to differ from those of the squarks and sleptons, the 
non-universal Higgs model or NUHM1 (NUHM2).
Other ``simplified'' versions of the MSSM that are based on (some)
unification at a higher scale are minimal Gauge mediated
SUSY-breaking (mGMSB) and minimal Anomaly mediated SUSY-breaking
(mAMSB). 

There have been many previous studies of the CMSSM parameter
space using EWPO, BPO and astrophysical 
data~\cite{ehow3,ehow4,ehoww,other,othercomp,mastercode1,mastercode2}, 
partially using Markov-chain Monte Carlo (MCMC) techniques. 
These analyses extracted the preferred values for the CMSSM parameters
They differ in the precision observables that have been considered,
the level of sophistication of the theory predictions that have been
used, and  the way the statistical analysis has been performed (e.g.\
Bayesian vs.\ Frequentist).
Many CMSSM analyses found evidence for a relatively low SUSY mass
scale, but no strong preference for any $\tb$ region has been
found. A mild preference for $\tb \approx 10$ in the stau-coannihilation
region was found in~\citeres{mastercode1,mastercode2,trotta}.
Deviations may arise from differences in the treatments of the 
theoretical constraints imposed by the $\br(b \to s \ga)$ and $(g-2)_\mu$
measurements, as analyzed in \citere{mastercode2}. 
A comparison of Bayesian analyses yielding varying results using
different priors was made in~\cite{othercomp}. The prior dependence can
be avoided by the use of a pure likelihood analysis.

More analyses have been performed focusing on other GUT based models
such as the NUHM2~\cite{nuhm2}, mGMSB and mAMSB~\cite{asbs3} or finite
models~\cite{fut}.


\section{DETERMINATION OF SUSY MASSES}

As outlined above, many$^2$ studies have been performed to determine the
SUSY mass scales using EWPO, BPO and astrophysical data. 
Here we will review two recently obtained
results~\cite{mastercode2,asbs3} that are based on a Frequentist
approach. However, they differ significantly in the choice of precision
observables and the statistical methods.

The study performed in \citere{asbs3} compared the models CMSSM, mGMSB
and mAMSB. A relatively small set of
precision observables, $\MW$, $\sweff$, $\amu$, $\Mh$, 
$\br(b \to s \ga)$ and $\br(B_s \to \mu^+\mu^-)$, was used to construct
a ``simple'' $\chi^2$ function. No CDM bounds have been taken into
account. 
The three models were scanned with $\sim 10^5$ points each, randomly
sampled over the respective parameter space. The results for the best
fit points are summarized in \refta{tab:chi2min}. It is interesting to
note that despite mAMSB has one parameter less, the minimum $\chi^2$
value is lower by $\sim 1.5-2$ compared to the CMSSM and mGMSB. A more
detailed analysis of this effect is given in \citere{asbs3}.

\begin{table}[htb!]
\begin{tabular}{c|ccc} 
\cline{2-4} \multicolumn{1}{c|}{}
& CMSSM & mGMSB & mAMSB \\
\hline
$\chi^2_{\rm min}$       & 4.6 &  5.1   &  2.9 \\ \hline
$\MW$                   & 1.7 &  2.1   &  0.6 \\ 
$\sweff$                & 0.1 &  0.0   &  0.8 \\ 
$(g-2)_\mu$              & 0.6 &  0.9   &  0.0 \\
$\br(b \to s \ga)$       & 1.1 &  2.0   & 1.5 \\ 
$\Mh$                    & 1.1 &  0.1   & 0.0 \\ 
$\br(B_s \to \mu\mu)$ [$10^{-8}$] 
                         & 4.5 & 3.2 & 0.4 \\ \hline
$\MA$ [GeV] (best-fit)   & 394 & 547 & 616 \\ 
$\tb$ (best-fit)         &  54 &  55 &   9 \\ 
$\mu$ [GeV] (best-fit)   & 588 & 810 & 604 \\
\hline
\end{tabular}
\caption{%
Minimum $\chi^2$ values for the CMSSM, mGMSB and mAMSB.
Also shown are the individual contributions for $\MW$, $\sweff$, 
$(g-2)_\mu$, $\br(b \to s \ga)$ and $\Mh$, as well as the value of
$\br(B_s \to \mu^+\mu^-)$. 
Shown in the last two rows are the best-fit values for the low-energy
parameters, $\MA$, $\tb$ and $\mu$.
}
\label{tab:chi2min}
\end{table}

Furthermore shown in the last three rows of \refta{tab:chi2min} are the
best-fit values of the $\cp$-odd Higgs boson mass $\MA$, $\tb$ and $\mu$.
They indicate that the heavy Higgs bosons
corresponding to the  best-fit parameter points of the CMSSM and mGMSB
would be accessible at the LHC, whereas they would escape the LHC
detection for mAMSB~\cite{HiggsLHCreach}.

The predictions in the three soft SUSY-breaking scenarios for two SUSY
mass scales are shown in \reffis{fig:mstau}, \ref{fig:mgluino}~\cite{asbs3}. 
In the first figure 
the mass of the lighter scalar tau is shown in the CMSSM (left), mGMSB
(middle), mAMSB (right) scenario for $\mu > 0$ with their respective
total $\chi^2$. The region with $\De\chi^2 := \chi^2 - \chi^2_{\rm min} < 1$
is medium shaded (green), the $\De\chi^2 < 4$ region is dark shaded
(red), and the $\De\chi^2 < 9$ region is light shaded (yellow). The rest
of the scanned parameter space is given in black shading.
The light $\tilde\tau$ has its best-fit values at very low masses, and even
the $\De\chi^2 < 4$ regions hardly exceed $\sim 500 \gev$ in mGMSB and mAMSB.
Therefore in these scenarios there are good prospects for the ILC(1000)
(i.e.\ with $\sqrt{s} = 1000 \gev$).
Also the LHC can be expected to  
cover large parts of the $\De\chi^2 < 4$ mass intervals.
In the CMSSM scenario, on the other hand, this region exceeds
$\sim 1 \tev$ such that only parts can be probed at the ILC(1000) and
the LHC.

The predictions of the gluino mass, $\mgl$, are shown in 
\reffi{fig:mgluino}. 
As before, the masses are shown in the CMSSM (left), mGMSB (middle) and
mAMSB (right) scenarios for $\mu > 0$ with their respective total $\chi^2$.
The color coding is as in \reffi{fig:mstau}. 
The gluino masses in the $\De\chi^2 < 4$ regions
range from a few hundred GeV up to about 
$3 \tev$ in mGMSB, exhausting the accessible range at the LHC. In the
other two scenarios the $\De\chi^2 < 4 $ regions end at $\sim 2 \tev$ (mAMSB)
and $\sim 2.5 \tev$ (CMSSM), making them more easily accessible at the LHC
than in the mGMSB scenario.

We now turn to the study performed in \citere{mastercode2}, focusing on
the CMSSM and the NUHM1. A large set of EWPO and BPO has been used to
constrain the 
model, see \citeres{mastercode1,mastercode2} for details. Also the CDM
bounds have been taken into account.
In this analysis an MCMC technique (see, e.g., \citere{mcmc} and
references therein) has been used to sample efficiently the CMSSM
parameter space. 
A global $\chi^2$ function was defined, which combines all calculations 
with experimental constraints:
\begin{align}
\chi^2 &= \sum^N_i \frac{(C_i - P_i)^2}{\sigma(C_i)^2 + \sigma(P_i)^2}
       + \sum_i \frac{(f^{\rm obs}_{{\rm SM}_i} 
                  - f^{\rm fit}_{{\rm SM}_i})^2}{\sigma(f_{{\rm SM}_i})^2} 
\end{align}
Here $N$ is the number of observables studied, $C_i$ represents an 
experimentally measured value (constraint) and each $P_i$ defines a CMSSM
parameter-dependent prediction for the corresponding constraint.  The three
SM parameters $f_{\rm SM} = \{\Delta\alpha_{\rm had}, m_t, m_Z\}$
are included as fit parameters and constrained to be within their current
experimental resolution $\sigma(f_{\rm SM})$.
With this a $\chi^2$ probability function was constructed, 
$P(\chi^2,N_{\rm dof})$. This accounts correctly for the number of degrees of
freedom, $N_{\rm dof}$, and thus represents a quantitative measure for the
quality-of-fit.  Hence $P(\chi^2,N_{\rm dof})$ can be used to estimate the
absolute probability with which the CMSSM describes the experimental
data.

\begin{figure}[htb!]
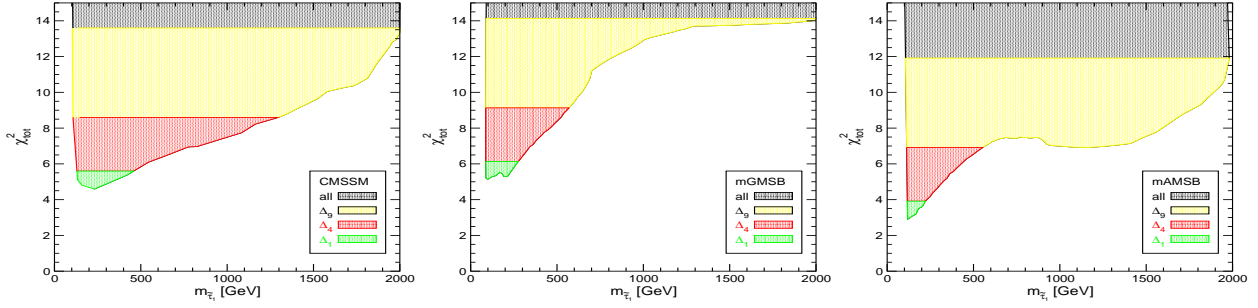

\includegraphics[width=0.33\textwidth,height=4cm]{asbs3_s_mass1a_cl}
\includegraphics[width=0.33\textwidth,height=4cm]{asbs3_g_mass1a_cl}
\includegraphics[width=0.33\textwidth,height=4cm]{asbs3_a_mass1a_cl}
\caption{
The mass of the lighter scalar tau is shown in the CMSSM (left), mGMSB
(middle), mAMSB (right) scenario for $\mu > 0$ with their respective
total $\chi^2$. The region with $\De\chi^2 := \chi^2 - \chi^2_{\rm min} < 1$
is medium shaded (green), the $\De\chi^2 < 4$ region is dark shaded
(red), and the $\De\chi^2 < 9$ region is light shaded (yellow). The rest
of the scanned parameter space is given in black shading~\cite{asbs3}.
}
\label{fig:mstau}
\end{figure}

\begin{figure}[htb!]
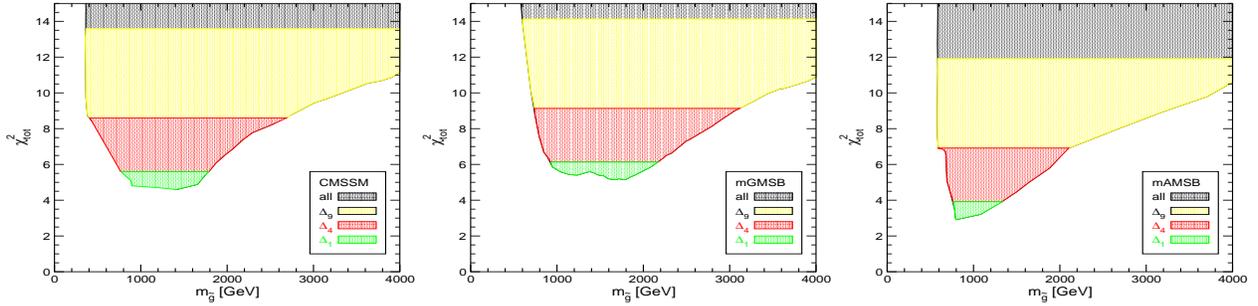

\includegraphics[width=0.33\textwidth,height=4cm]{asbs3_s_mass14_cl}
\includegraphics[width=0.33\textwidth,height=4cm]{asbs3_g_mass14_cl}
\includegraphics[width=0.33\textwidth,height=4cm]{asbs3_a_mass14_cl}
\caption{
The mass of the gluino is shown in the CMSSM (left), mGMSB
(middle), mAMSB (right) scenario for $\mu > 0$ with their respective
total $\chi^2$~\cite{asbs3}. The color coding is as in \reffi{fig:mstau}.
}
\label{fig:mgluino}
\end{figure}

\begin{figure}[htb!]
\includegraphics[width=0.46\textwidth]{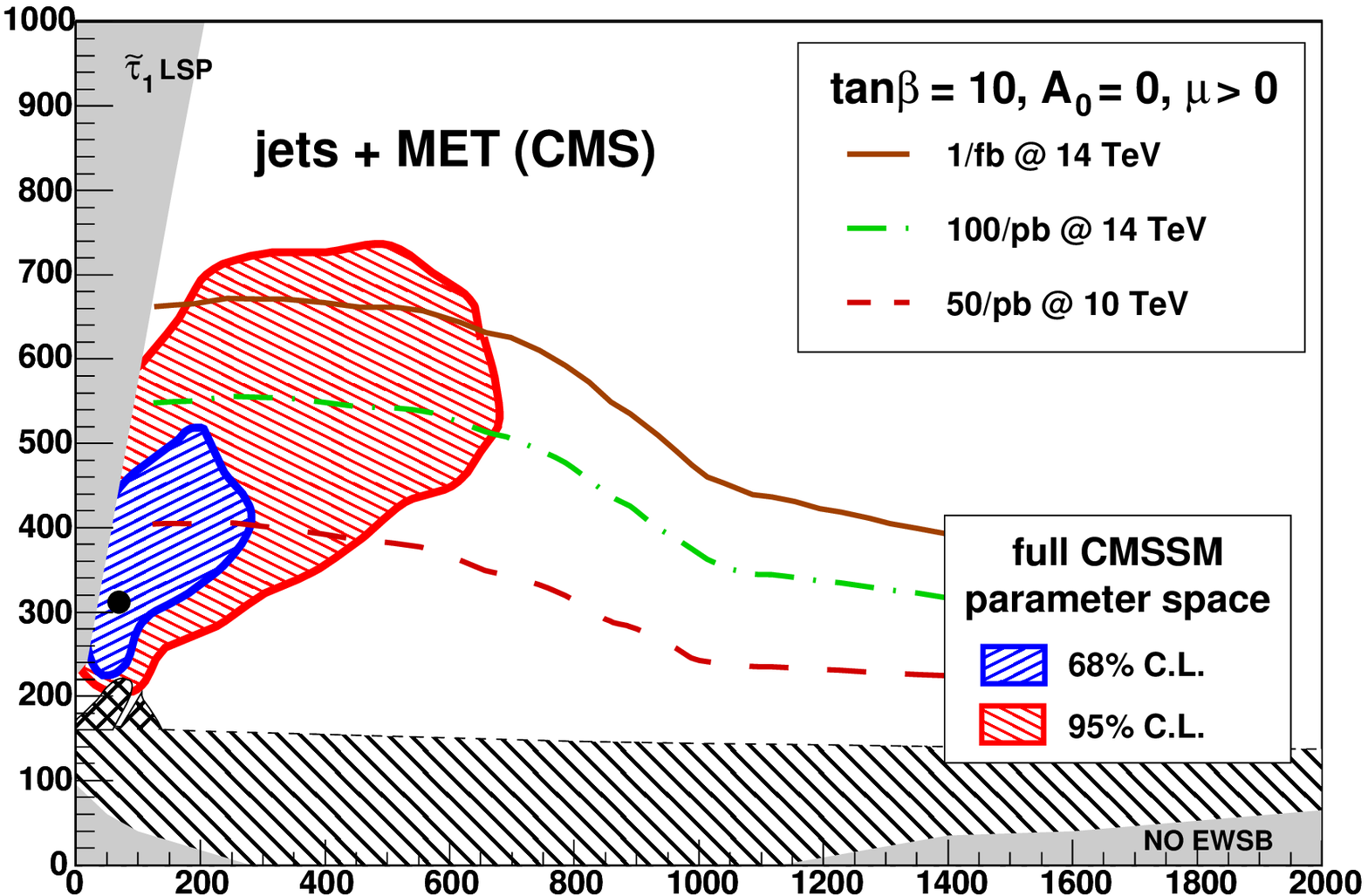}~~
\includegraphics[width=0.46\textwidth]{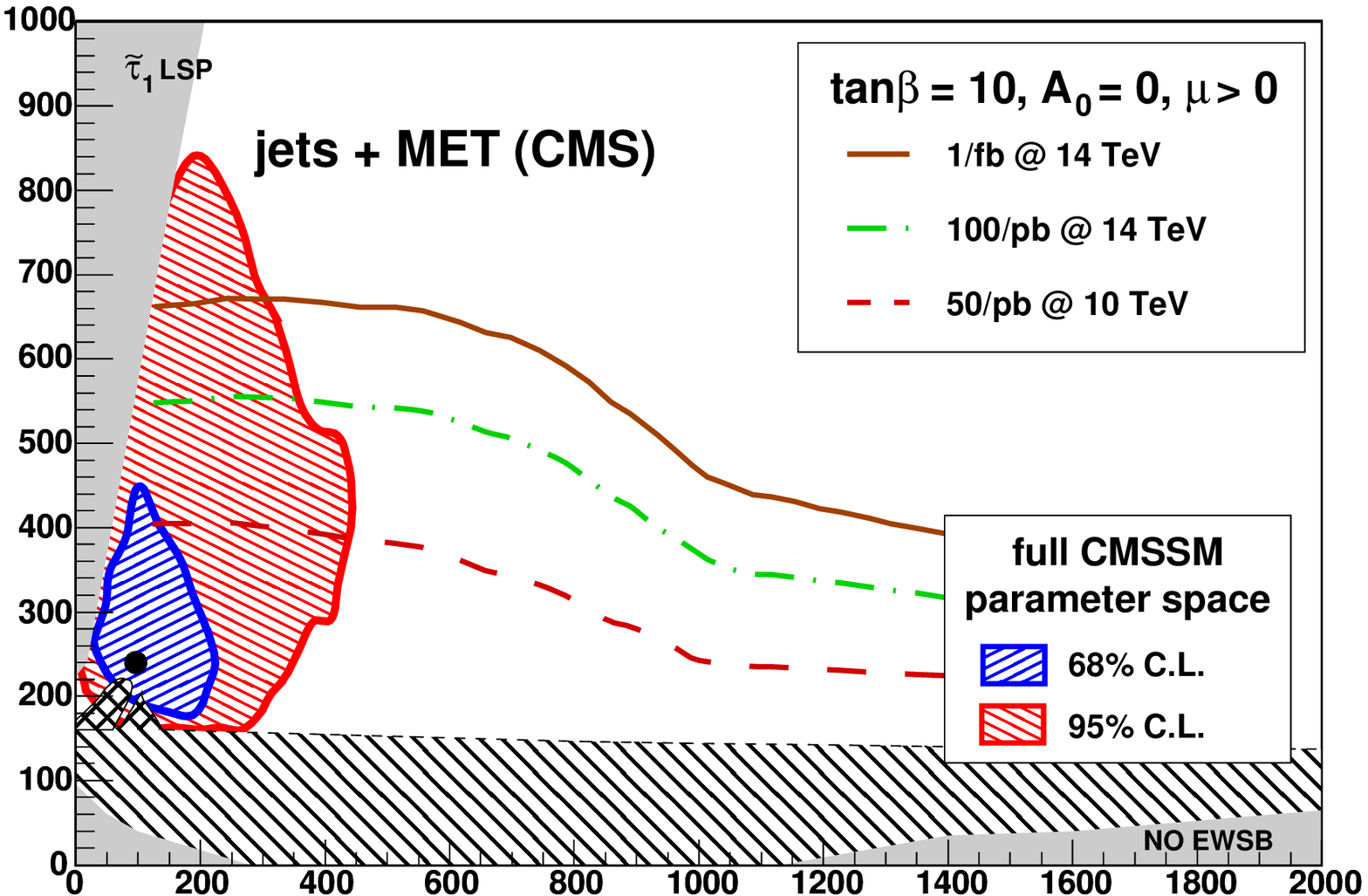}
\begin{picture}(0,0)
  \put(-265,-8){$m_0$ [GeV]}
  \put(-450,95){\begin{rotate}{90}$m_{1/2}$ [GeV]\end{rotate}}
  \put(-35,-8){$m_0$ [GeV]}
  \put(-220,95){\begin{rotate}{90}$m_{1/2}$ [GeV]\end{rotate}}
\end{picture}
\vspace{-0.5em}
\caption {The $m_0$--$m_{1/2}$ plane in the CMSSM (left) and NUHM1 (right). 
  The best-fit point is indicated by a filled circle, and the 
  68 (95)\%~C.L.\ contours from the fit as dark grey/blue (light
  grey/red) overlays, scanned over all $\tan\beta$ and $A_0$ values.
  The plot shows the $5\,\sigma$ discovery contours for jet + missing
  $E_T$ events at CMS with 1~fb$^{-1}$ at 14~TeV, 100~pb$^{-1}$ at 14~TeV and
  50~pb$^{-1}$ at 10~TeV centre-of-mass energy (for $\tan\beta = 10$ 
  and $A_0 = 0$).
} 
\label{fig:mc2}
\end{figure}

In \reffi{fig:mc2} we show the prediction derived in
\citere{mastercode2} for the CMSSM (left) and NUHM1 (right) parameters
based on the fit of the EWPO, BPO and astrophysical data.
The $m_0$--$m_{1/2}$ plane in the two scenarios is shown, where
the best-fit point is indicated by a filled circle, and the 
68 (95)\%~C.L.\ contours from the fit as dark grey/blue (light
grey/red) overlays, scanned over all $\tan\beta$ and $A_0$ values.
The CMSSM best-fit point has the parameters
$m_0 = 60$~GeV, $m_{1/2} = 310$~GeV, $A_0 = 240$~GeV and $\tan\beta = 11$,
the NUHM1 best-fit point has $m_0 = 100$~GeV, $m_{1/2} = 240$~GeV, 
$A_0 = -930$~GeV, $\tan \beta = 7$, $m_H^2 = -6.9 \times 10^5$~GeV$^2$.
Furthermore shown are the $5\,\sigma$ discovery contours for jet + missing
$E_T$ events at CMS with 1~fb$^{-1}$ at 14~TeV, 100~pb$^{-1}$ at 14~TeV and
50~pb$^{-1}$ at 10~TeV centre-of-mass energy. They have been obtained
for $\tb = 10$ and $A_0 = 0$, but do not vary significantly with these
two parameters. 
The dark shaded area in \reffi{fig:mc2} at low $m_0$ and high $m_{1/2}$
is excluded due to a scalar tau LSP, the light shaded areas at low
$m_{1/2}$ do not exhibit electroweak symmetry breaking. The nearly
horizontal line at $m_{1/2} \approx 160$~GeV has 
$m_{\tilde \chi_1^\pm} = 103$~GeV, and the area below is excluded by LEP
searches. Just above this contour at low $m_0$ in the lower panel is the
region that is excluded by trilepton searches at the Tevatron.
One can see that the for the CMSSM as well as for the NUHM1 the 
68\% likelihood contour is 
well covered by the 14~TeV/100~pb$^{-1}$ discovery reach, and even
the 10~TeV/50~pb$^{-1}$ reach would be sufficient to discover SUSY at
the best-fit point. This offers very good prospects for the initial LHC
running.


\section{CONSTRAINING $\Mh$}

Since the lightest MSSM Higgs boson mass can be predicted in terms of
the other model parameters (see \citeres{PomssmRep,HiggsReviews} for
reviews), its preferred value can be fitted to the EWPO (and BPO) data,
similar to the ``Blue Band'' plot in the SM~\cite{LEPEWWG}. 
For these analyses it is crucial {\em not} to include the $\chi^2$
contribution of $\Mh$ itself into the fit.

\begin{figure}[htb!]
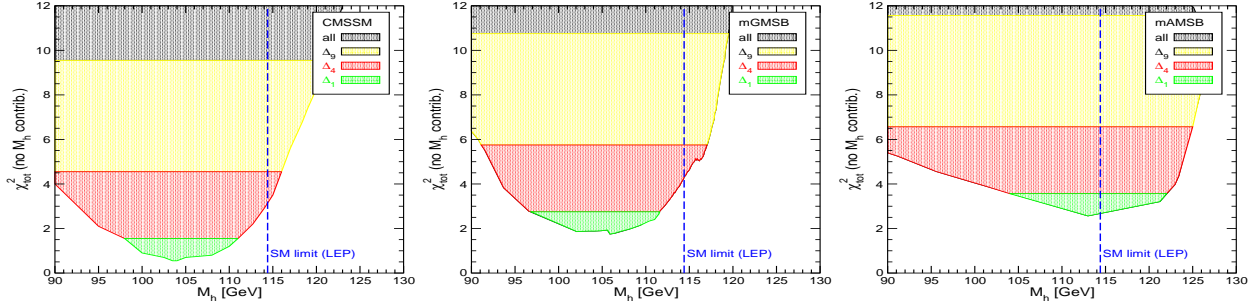

\includegraphics[width=0.33\textwidth,height=4cm]{asbs3_s_mass02_cl}
\includegraphics[width=0.33\textwidth,height=4cm]{asbs3_g_mass02_cl}
\includegraphics[width=0.33\textwidth,height=4cm]{asbs3_a_mass02_cl}
\caption{
The $\Mh$ values in the CMSSM (left), mGMSB (middle) and mAMSB (right)
scenarios for $\mu > 0$ with their respective $\chi^2$, where the $\chi^2$
contribution of the $\Mh$ itself has been left out. 
The color coding is as in \reffi{fig:mstau}. The SM limit of 
$114.4 \gev$ obtained at LEP is indicated with a dashed (blue) line.
}
\label{fig:mh}
\end{figure}

\begin{figure*}[htb!]
\begin{picture}(500,190) 
  \put(20,-10){ \resizebox{7.0cm}{!}
             {\includegraphics{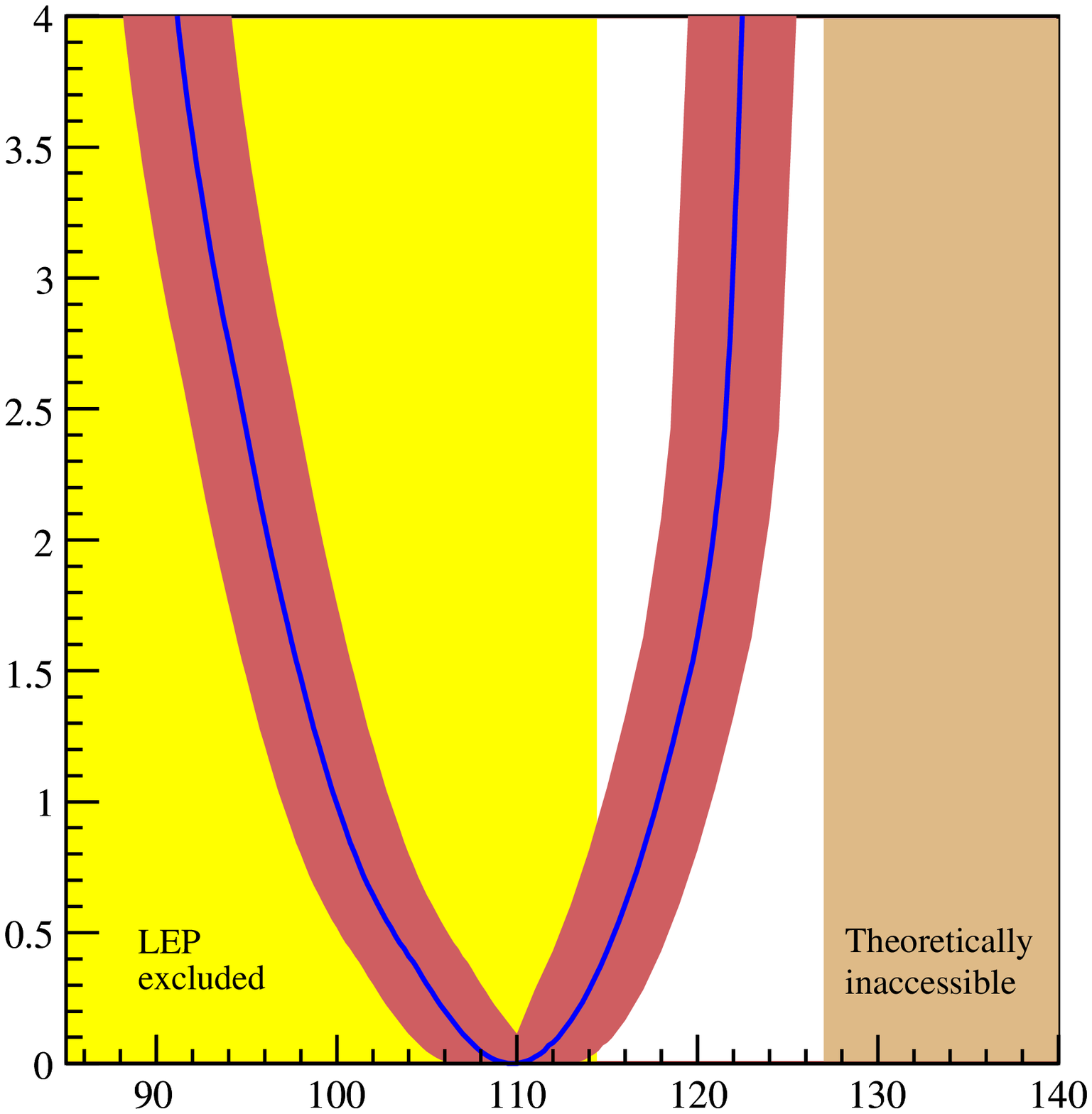}}  }
  \put(85, 140){CMSSM}
  \put(170, -7){$\Mh$ [GeV]}
  \put(20, 150){\begin{rotate}{90}$\Delta \chi^2$\end{rotate}}
  \put(250,-10){ \resizebox{7.0cm}{!}
               {\includegraphics{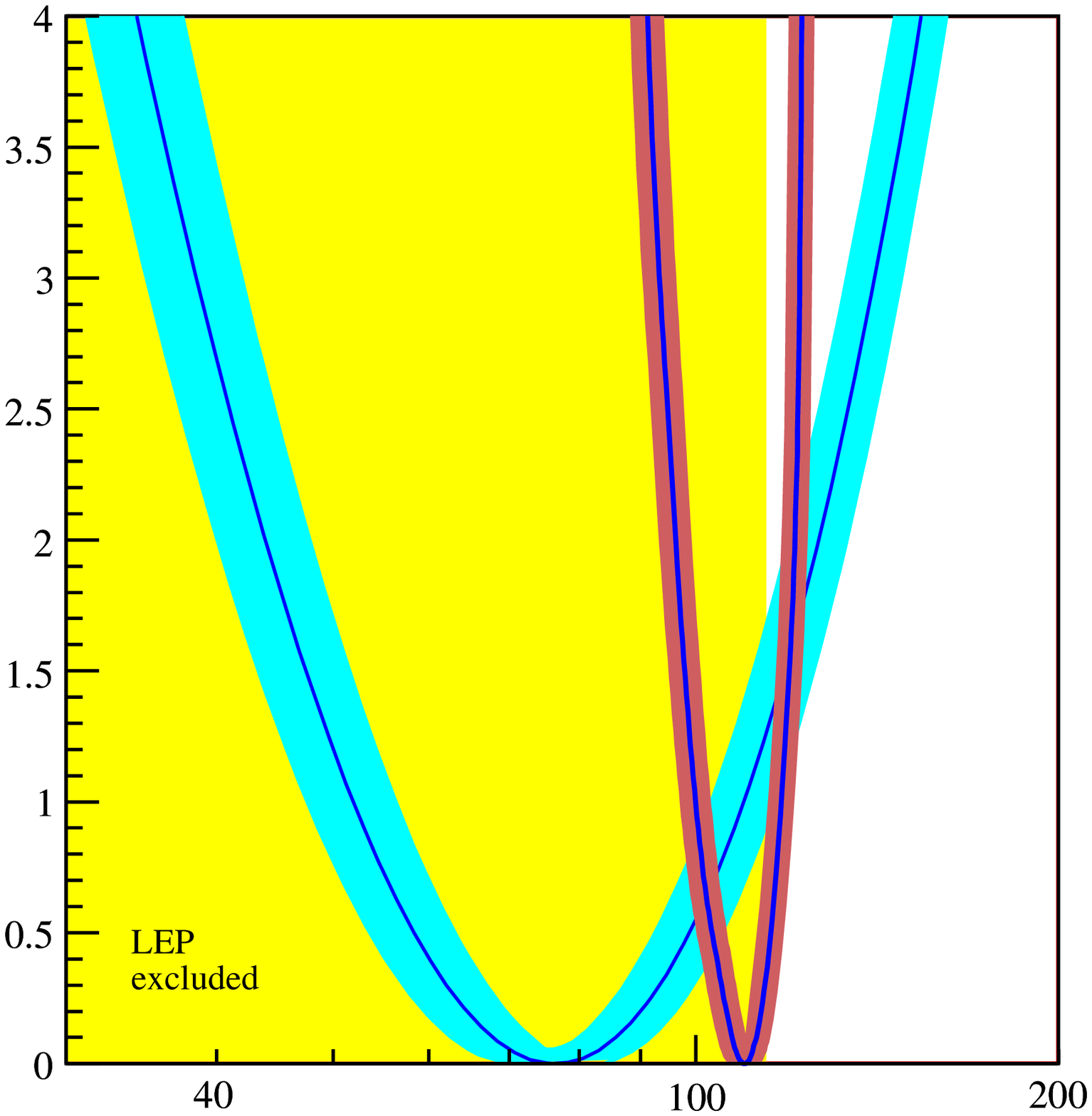}}  }
  \put(375, -7){$M_{\rm Higgs}$ [GeV]}
  \put(255, 150){\begin{rotate}{90}$\Delta \chi^2$\end{rotate}}
  \put(390, 30){CMSSM}
  \put(310, 140){SM}
\end{picture}
\caption{Left: Scan of the lightest Higgs boson mass versus $\De\chi^2$.
  The curve is the result of a CMSSM fit
  using all of the available constraints (see text). The direct limit
  on $\Mh$ from LEP~\cite{LEPHiggsSM,LEPHiggsMSSM} is not included.
  The red (dark gray) band represents the total
  theoretical uncertainty from unknown higher-order corrections,
  and the dark shaded area on the right above $127 \gev$ 
  is theoretically inaccessible (see text).
  Right:  Scan of the Higgs boson
  mass versus $\Delta \chi^2$ for the SM (blue/light gray), as
  determined by \cite{LEPEWWG} using all available electroweak
  constraints, and for comparison, with the CMSSM scan superimposed
  (red/dark gray). 
}
\label{fig:mh_vs_chi2}
\end{figure*}

We first review the results obtained in \citere{asbs3}, where the three
soft-SUSY breaking scenarios CMSSM, mGMSB and mAMSB are compared. As
described above, the CDM has not been considered here.
In \reffi{fig:mh}. 
$\Mh$ is shown in the CMSSM (left), mGMSB (middle) and mAMSB (right)
scenarios for $\mu > 0$ with the corresponding $\chi^2$, where the $\chi^2$
contribution of $\Mh$ itself has been left out. In this way the plot
shows the indirect predictions for $\Mh$ without imposing the bounds
from the Higgs boson searches at LEP~\cite{LEPHiggsSM,LEPHiggsMSSM}.
In all three scenarios
a shallow minimum can be observed. The $\De\chi^2 < 1$ regions are in the
intervals of $\Mh = 98 \ldots 111 \gev$ (CMSSM), 
$97 \ldots 112 \gev$ (mGMSB) and $104 \ldots 122 \gev$ (mAMSB). In all
three scenarios the $\De_4$ regions extend beyond the LEP limit of 
$\Mh > 114.4 \gev$ at the 95\% C.L. shown as dashed (blue) line in
\reffi{fig:mh} (which is valid for the three soft
SUSY-breaking scenarios, see \citeres{asbs1,ehow1}). 
All three scenarios have a
significant part of the parameter space with a relatively low total $\chi^2$
that is in agreement with the bounds from Higgs-boson searches at LEP.
Especially within the mAMSB scenario the $\De\chi^2 < 1$ region extends
beyond the LEP bound of $114.4 \gev$.

The fact that the minimum in \reffi{fig:mh} is relatively 
sharply defined is a general consequence of the MSSM, where the
neutral Higgs boson mass is not a free parameter.
The theoretical upper bound $\Mh \lsim 125 \gev$ (depending somewhat on
the GUT scenario)
explains the sharper rise of the $\De\chi^2$ at large $\Mh$
values. In the SM, $\MHSM$ is a
free parameter and only enters (at leading order) logarithmically in
the prediction of the precision observables. In the MSSM this
logarithmic dependence is still present, but in addition $\Mh$
depends on $\mt$ and the SUSY parameters, mainly from the scalar
top sector. The low-energy SUSY parameters in turn are all connected
via RGEs to the GUT scale parameters.  
The sensitivity on $\Mh$ in the analysis of \citere{asbs3} (and also in
\citere{mastercode1}, see below) is therefore the combination of the indirect
constraints on the few free GUT parameters and $\tb$ and the fact that
$\Mh$ is directly predicted in terms of these parameters.

A fit as close as possible to the SM fit for $\MHSM$ 
has been performed in \citere{mastercode1}. 
All EWPO as in the SM~\cite{LEPEWWG} (except $\Ga_W$, which has a
minor impact) were included, supplemented by the CDM constraint, 
the $(g-2)_\mu$ results and the $\br(b \to s \ga)$ constraint.
The top quark mass used in this fit was $\mt = 170.9 \pm 1.8 \gev$.
The $\chi^2$ is minimized with respect to all CMSSM parameters for
each point of this scan. Therefore, $\De \chi^2=1$ represents the
68\% confidence level uncertainty on $\Mh$.
Since the direct Higgs boson search limit from LEP is not used in this
scan the lower bound on $\Mh$ arises as a consequence of 
{\em indirect} constraints only, as in the SM fit.

In the left plot of \reffi{fig:mh_vs_chi2}~\cite{mastercode1} the
$\De\chi^2$ is shown as a function of $\Mh$ in the CMSSM. The area
with $\Mh \ge 127$ is theoretically inaccessible. The right
plot of \reffi{fig:mh_vs_chi2} shows the red band parabola from the
CMSSM in comparison with the blue band parabola from the SM. There is a
well defined minimum in the red band parabola, leading to a prediction
of~\cite{mastercode1}  
$\Mh^{\rm CMSSM} = 110^{+8}_{-10}\;{\small{\rm(exp)}} 
                     \pm 3\;{\small{\rm(theo)}\gev ,}$
where the first, asymmetric uncertainties are experimental and the
second uncertainty is theoretical (from the unknown higher-order
corrections to $\Mh$~\cite{mhiggsAEC,PomssmRep}). 
The most striking feature is that even {\em without} the direct
experimental lower limit from LEP of $114.4 \gev$
the CMSSM prefers a Higgs boson mass which is
quite close to and compatible with this bound.



\begin{theacknowledgments}
We thank the authors of \citeres{mastercode2,asbs3} with whom the
results presented here have been derived.
\end{theacknowledgments}


\bibliographystyle{aipproc}   


}

\end{document}